\documentstyle[preprint,aps,pra]{revtex}
\begin{document}
\preprint{IASSNS-96/33}
\draft

\title{Complexity, Tunneling and Geometrical Symmetry}
\author{L. P. Horwitz$^{\text{(1)(2)(3)}}$, J. Levitan$^{\text{(2)(4)}}$
and Y. Ashkenazy$^{\text{(2)}}$}
\address{(1) School of Natural Sciences, Institute for Advanced Study,
Princeton, N.J. 08540\\
(2) Department of Physics, Bar-Ilan University, Ramat-Gan 52900,
Israel\\
(3) School of Physics, Raymond and Beverly Sackler Faculty of Exact Sciences\\
Tel-Aviv University, Ramat-Aviv, Israel.\\
(4) The Research Institute, The College of Judea and Samaria,\\
Kedumim and Ariel, P.O.B. 3 Ariel, 44837 Israel.}
\date{\today}
\maketitle

%\twocolumn[
%\date{\today}
%\maketitle
%\widetext

%\vspace*{-1.0truecm}

\begin{abstract}
{
It is demonstrated in the context of the simple one-dimensional example
of a barrier in an infinite well, that highly  complex behavior of
the time evolution of a wave function is associated with the almost
degeneracy of levels in the process of tunneling. Degenerate conditions are
obtained by shifting the position of the barrier. The complexity strength
depends on the number of almost degenerate levels which depend on geometrical
symmetry.
The presence of complex behavior is
studied to establish correlation with spectral degeneracy.
}
\end{abstract}

\pacs{
\hspace{1.9cm}
}
%]
\narrowtext

\newpage

Tunneling processes have become of considerable interest as one of
the possible mechanisms for creating highly complex behavior in the
structure of the quantum wave function.
Tomsovic and Ullmo \cite{Tomso94}
found that there is an interesting correlation between classical chaotic
behavior and the rate of tunneling in the corresponding quantum system.
The conclusion of their study is that chaos facilitates tunneling.

On the other hand, Pattanayak and Schieve \cite{PatanPRL}
\cite{PatanPRE} found chaotic
behavior in the semi-classical phase space (defined by expectation value)
of a one-dimensional time-independent Duffing oscillator where new
variables, associated with dispersion of the quantum states, are defined
and included in the description of the system. They concluded that
quantum tunneling plays a crucial role for the chaotic behavior in the
corresponding semi-classical maps. They have argued \cite{PatanPRE}
that the spectrum
becomes more complicated in the neighborhood of the separatrix.

In a recent study, we have considered a model in which tunneling leads to
highly complex behavior of the quantum wave function and its time dependence.
The spectrum, as anticipated by Pattanayak and Schieve \cite{PatanPRL}
\cite{PatanPRE}, indeed
makes a transition to more complex behavior in the presence of a
classical separatrix \cite{Berko}. It is clear that the almost degeneracy of
levels is necessary for the
existence of significant tunneling. We directly investigate, in this work, the
effect of almost degeneracy in the presence of tunneling on the complexity and
the behavior of the development of the wave function. This criterion is,
in fact,  closely analogous to the criterion of overlapping resonances
for the onset of classical chaos \cite{Dana}.

The model we shall use is related to the one we previously explored \cite{Ash},
i.e., a barrier embedded in an infinite well. By displacing the barrier in
the double well system (to the right or to the left),
certain positions are passed
where the system becomes strongly almost-degenerate.
These positions occur at almost
commensurate intervals.  It is exactly for those positions that one
may find significant
tunneling accompanied by complex behavior.
We show, moreover, that in the cases of very high degeneracy, tunneling from
left to right has exponential decay, on a significant interval of time, but
at other positions, where almost degenerate conditions are somewhat weaker,
the transition curve develops strong oscillations.

In a study by Nieto {\it et al.} \cite{Nieto}, it was shown
 that tunneling
in an asymmetric double well is a very sensitive function of the potential.
 The behavior of the development
of the wave function under these conditions was not, however,
discussed there.

The calculation in this work is done for a square barrier of height $V=5$ and
width $w=x_2-x_1=2$ ($x_1$ is the left boundary of the barrier and $x_2$ is
the right boundary; we
 take $\hbar \equiv 1,2m\equiv 1$) embedded in an infinite
well of width $2l=110$ (interval $(-l,l)$). In the calculation, an exact
analytic expression is evaluated on the computer; there is no accumulation
of error for large times, since there is no integration over time.

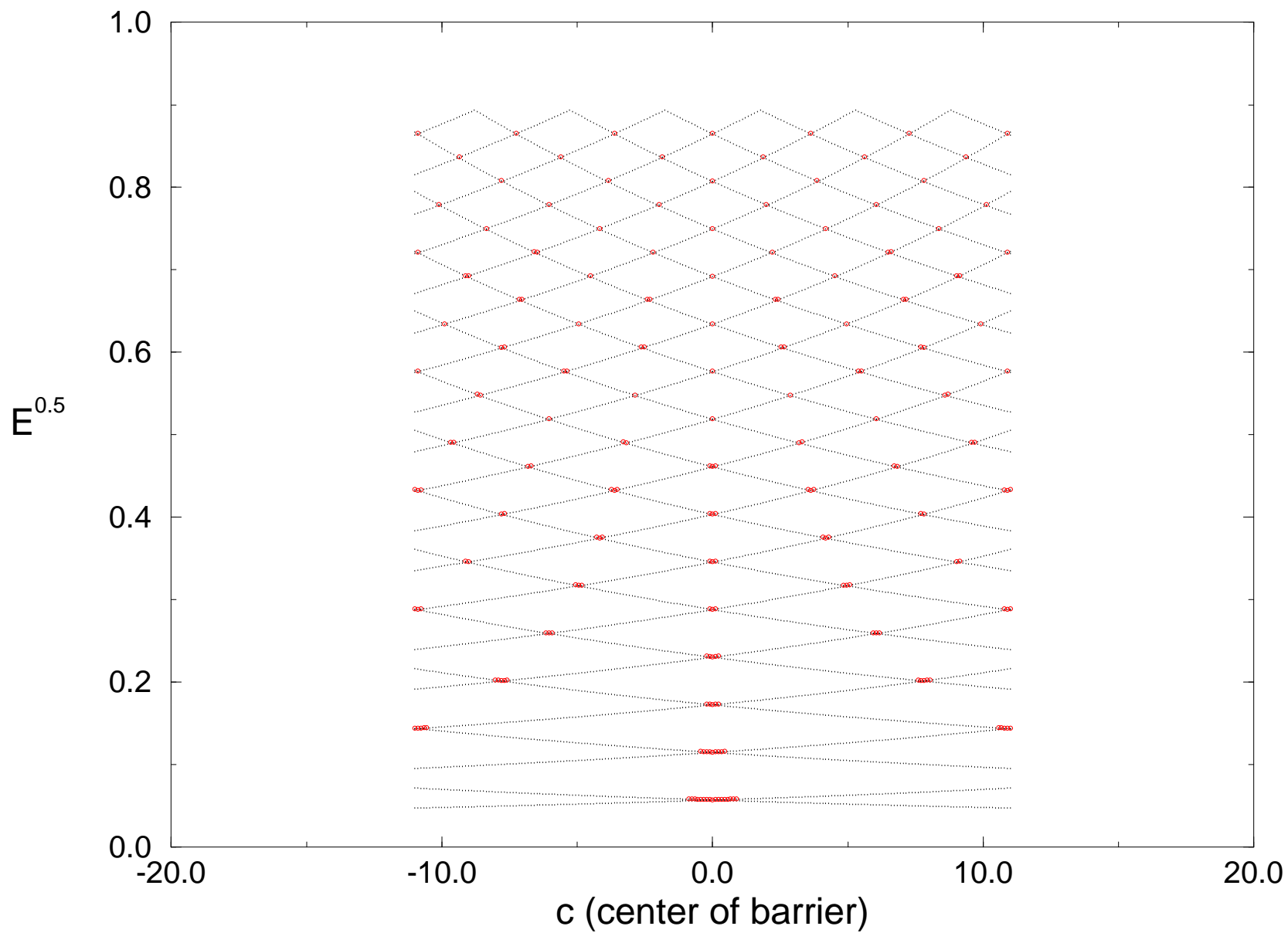
\begin{figure}[tbp]
\caption{ Square root of energy eigenvalues as a function of the center of the
barrier. }
\label{fig.1}
\end{figure}

We first discuss the energy spectrum according to the location
of the barrier. In Fig. \ref{fig.1}
 one can see the lower energy levels ``almost
crossing'' (the levels do not cross, but can become very close to each other)
at several locations of the barrier.
In the middle (the position $c$ of the center of the barrier is taken at
$c=0$; generally, $c={1 \over 2}(x_1+x_2)$)
every energy level is almost degenerate. In other discrete
locations, one finds almost degeneracy for every second level,
every third level, and so on.

As an approximation to our model, consider two
separate infinite wells with widths $l+x_1$ and $l-x_2$. The
energy levels of the two separate wells exhibit very similar behavior
to that of the finite barrier,
but in this case, exact degeneracy occurs, according to the
geometrical configuration of the system.
For $x_2=-x_1$, we have complete symmetry and all levels are degenerate.
When the width of the left well is twice the
width of the right well or vice versa,
one obtains degeneracy for every second energy level
(of three levels, two are degenerate).
If the width of the left well is one third of that of the
right well, every third level is degenerate,
and so on.
This follows from the relations
$E_{n_l} = \hbar ^2(\pi n_l)^2/2m(l+x_1)^2$ and
$E_{n_r} = \hbar ^2(\pi n_r)^2/2m(l-x_2)^2$; if
$n_l/n_r = (l+x_1)/(l-x_2)$, then $E_{n_l} = E_{n_r}$
($n_l$ and $n_r$ are positive integers). The locations for which these
degeneracies occur are :
$c = [(n_r - n_l)/(n_r + n_l)](l - w/2 )$.
In every $n_l + n_r$ levels we have at least one degeneracy.
In our model (barrier in infinite well),
 the behavior
is very similar to the problem of separate wells
for the lower energy levels (for the higher energy levels
there are, in fact, no crossings); degeneracy locations are, moreover,
shifted slightly forwards the center.

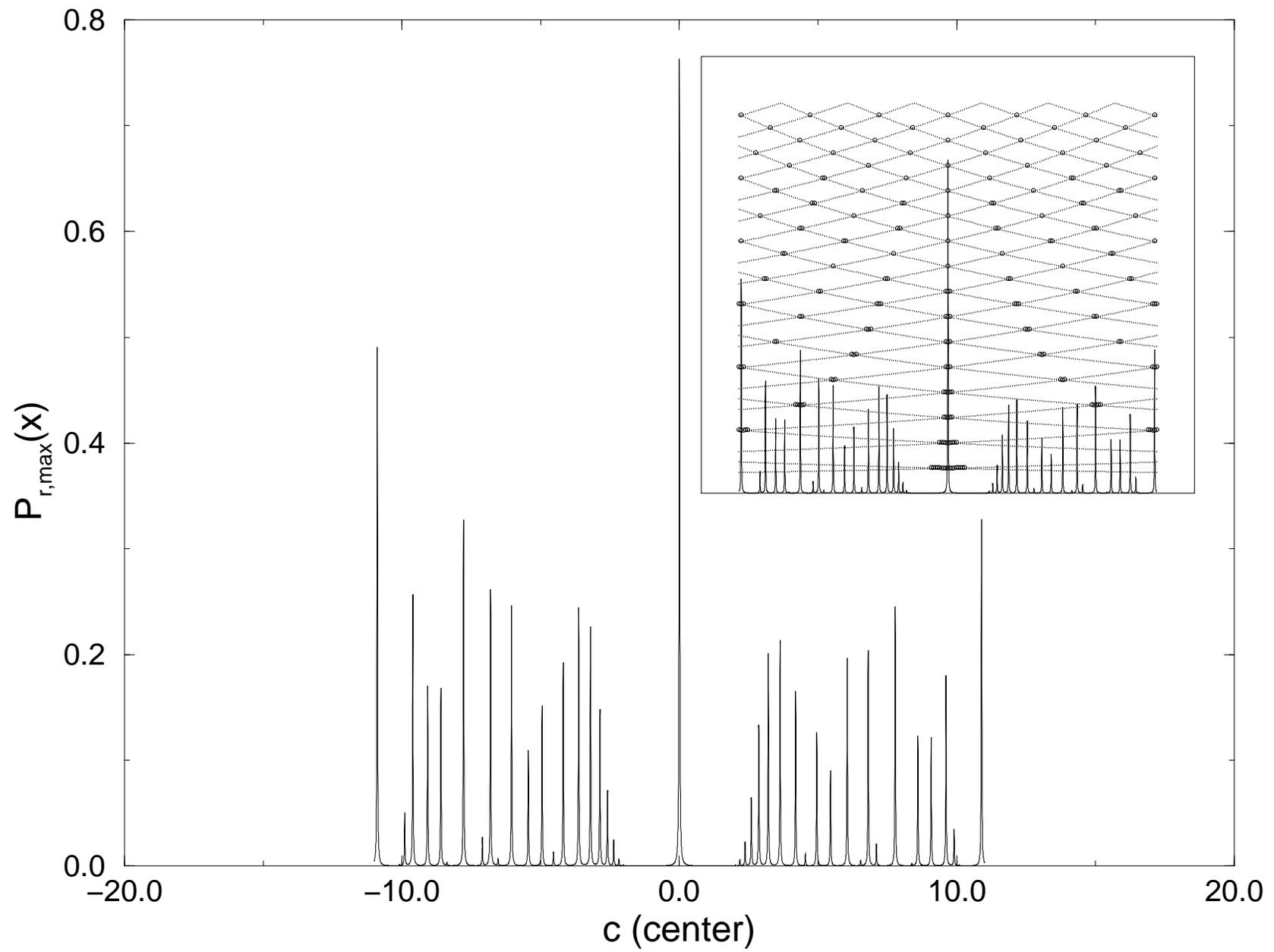
\begin{figure}[tbp]
\caption{ Maximum probability to be in right side of the barrier.
The inset shows partial correspondence between RE (resonance enhancement)
and almost degenerate levels. }
\label{fig.2}
\end{figure}

In order to study the tunneling process, we constructed a wave packet
 approximating the  form
$\psi (x,0) = c \exp ( -(x-x_0)^2 / (4 \sigma ^2) + i k_x ) $
(where $\sigma = 5$, $k_x \sim 0.45$ and $x_0 = (x_1-l)/2)$,
from 28 or less of the first energy levels, located
in the middle of the left side of the barrier (the normalization of the wave
packet is approximately 0.9999). The average energy is approximately
$0.1 \times V$.
We then measured the maximum probability
to be on right side of the barrier during a very long time interval
($t_{max} \sim 2 \times 5 \times 10^5 $).
The calculation is done for the
central region of the well (from $-l/5$ to $l/5$) to avoid
a phase space imbalancing effect
(if the barrier is located too close to the left side, for example, the
probability to be in the left side is much smaller than the probability to
be in the right side, just because the available space is much less).
It can be seen
clearly from Fig. \ref{fig.2}, that
just several positions (which we call RE, for
resonance enhancement) allow tunneling, while in most
regions the wave packet is trapped in the left side. The strongest RE is
found in the center of the well where we have complete symmetry.
Fig. \ref{fig.1}
and Fig. \ref{fig.2} show complete correspondence;
the strength of the RE's depend
on the number of degenerate energy levels (Fig.\ref{fig.2}, inset).
The second strongest RE in the picture
is found where the system has
almost degeneracy for every fifth level ($c \sim -54/5$ and $c \sim 54/5$).

An additional factor that one must consider is the projection
of the initial state (Gaussian wave packet) on different eigenfunctions,
i.e., the coefficients of the representation.
It is clear that the strength of an  RE depends on the number of almost
degenerate states that have a large overlap with the initial state.
Thus, in order to get strong RE, there must exist at least one pair of
eigenstates, $|j-1>$, $|j>$, which fulfill two conditions : 1) Almost
degeneracy of levels ($E_{j-1} \sim E_j$),
 and, 2) the scalar product of the eigenstate
with the initial state is large enough (i.e.,
in our case, $(\psi (0) , \phi_j )$ {\rm  appreciable compared to unity}).
Condition 1 forms a general underlying
 symmetry of the system, while condition 2 is a requirement for the symmetry
effects to be realized.

In fact, almost degeneracy of levels (condition 1), $E_{j-1} \sim E_{j}$,
implies symmetry properties of a pair of eigenstates, $|j-1>$ and $|j>$.
Let us denote the part of the eigenfunction on the left side of the barrier as
$\phi_L$, and on the right side of the barrier as
$\phi_R$ (for simplicity we neglect the function under the barrier, since
the eigenfunctions are small in this region). Almost degeneracy of
levels and orthogonality implies that the
eigenfunctions are almost the same on one side the barrier, and opposite in
sign on the other side of the barrier, i.e., $\phi_{L,j-1} \sim
\phi_{L,j} \equiv \phi_L$ and $\phi_{R,j-1} \sim -\phi_{R,j} \equiv -\phi_R$,
or vice versa. Moreover, the eigenvalue condition requires that :
$\int _{L} |\phi_L|^2 \sim \int _{R} |\phi_R|^2$, while the normalization
condition requires : $\int _{L} |\phi_L|^2 + \int _{R} |\phi_R|^2 \sim 1$.
Thus, $\int _{L} |\phi_L|^2 \sim \int _{R} |\phi_R|^2 \sim {1 \over 2} $.
These properties implies symmetry and
antisymmetry in the central position.
We conjecture that this symmetry is the essential property of the
central position ($c=0$).

The wave function for the main, central
RE exhibits a complex behavior for the evolution. This complexity is
due to the large
number of almost degenerate levels. When we measure a physical quantity,
the difference between levels determines the time dependence
(the time dependent phase is computed according to
$\Delta E_{ij}$). The very small frequency due
to almost degeneracy implies a very large recurrence time. On the other hand,
large energy differences are associated with short time scales.
The influence of these types of frequencies
can be seen in most of the results.
However, this behavior does not occur for the total
probability in the left side of the well
as a function of time, as seen in Fig. \ref{fig.3}a.
The curves are smooth and do not
reflect the influence of the short time scale.
As explained before,
at RE locations there exist, at least, one pair of eigenfunctions
$\phi_{j-1}$, $\phi_j$ which are approximately
the same on the left side and
$\int_{-l}^{x_1} \phi _{j-1} \phi _{j}$ is appreciable (approximately $1 \over
2$). The influence of
other (non-neighboring) eigenfunctions on $P_{left}$, tend to be
small. The result is
a combination of periodic functions (the number of these functions corresponds
to the number of pairs that fulfil the two conditions for RE),
with very small frequency differences.

The transition
from one side to another, when $c=c_0=0$, exhibits approximate exponential
behavior for times not too short or too long (Fig. \ref{fig.3}a).
We observe similar behavior, but less clear, in other locations
that have very strong RE (locations such as, $c \sim -54/3$ and
$c \sim -54/2$).
It appears that this exponential decay is due to the behavior of a sum
of periodic functions with very small different frequencies, as can be seen
in Fig. \ref{fig.3}a.
After this interval of decay, $P_{left}$ enters a domain of large oscillations.

In Fig. \ref{fig.3}a, we show also the results of
the same calculation for some other RE locations. As expected,
we find a periodic (or almost periodic) behavior. The Fourier transform
(Fig. \ref{fig.3}b) shows very strong
frequency peaks that fit to the most dominant
almost degenerate energy levels,
and show clearly that very small frequencies dominate the motion.

The almost
degeneracy of levels that produces a high level of complexity
i.e., chaotic-like behavior, occurs in the presence of strong tunneling.
We have observed this effect in ref. \cite{Ash}.
We study here one of the most clear ways to display this connection.
We compare, in Fig. \ref{fig.3}a, results of four positions. For
$c_1=-10.89$, there is a large RE with
three pairs of dominant almost degenerate eigenstates, and for
 $c_2=-3.63$ there is a smaller RE with one pair of dominant almost
degenerate eigenstates. The choice of $c_3=-1$ results in no RE, no
 degeneracy and
no tunneling at all. The main RE, at $c=c_0=0$,
reflects a very complex behavior as we have shown in ref. \cite{Ash}.

In Fig. \ref{fig.4} the entropy defined by
$ S(t) = - \int | \psi (x,t) |^2 \ln | \psi (x,t) |^2 dx $ is computed.
For $c_0$, shown in Fig. \ref{fig.4}a, the entropy rises sharply
accompanied by high frequency
oscillations
and then remains a long time in a
``quasi-equilibrium state''.
The second RE, $c_1$, shown in Fig. \ref{fig.4}b,  shows a tendency to
 recurrence
after 280000 time units, while $c_2$, shown in Fig. \ref{fig.4}c,
returns to almost the initial condition after
approximately 80000 time units. The entropy for non-RE locations shows
almost periodic behavior (Fig. \ref{fig.4}d; the inset shows the
structure at increased scale).
A similar behavior can be seen for $\rho(t) = < <x^2> - <x>^2 >$.

Comparison between different locations shows that one can characterize the
behavior by two time scales. The first is the time
of approach to equilibrium ($\Delta t_{eq}$),
and the second is the recurrence time ($\Delta t_{rc}$).
The approach to equilibrium time corresponds
to averaging small frequencies (i.e.,
$< \Delta E >$ over all energies that satisfy almost degeneracy).
One can  easily identify $\Delta t_{eq}$ from Fig. \ref{fig.3}a
 (for $c_0$) and from
Fig. \ref{fig.4}a, while $\Delta t_{rc}$ can be calculated
analytically from the known eigenvalues.
It appears
that ${{\Delta t_{eq}} \over {\Delta t_{rc}}}$ can give a measure for
the complexity of the behavior of the system, as seen in these results.
In all cases the, $\Delta t_{eq}$, is approximately the same, while
$\Delta t_{rc}$ is changed drastically from $c_0$ to $c_3$ (one can not
recognize recurrence in the $c_0$ location, while $c_1$ and $c_2$
exhibit almost recurrence as mentioned in previous paragraph). Thus, the
ratio of these two time scales is largest for $c_3$ (largest RE and
maximum complexity), and decreases when the RE's become stronger (or when
the complexity becomes stronger).

In Fig. \ref{fig.5}, we compare the $ <p> : <x> $ maps, sampled at peak times
(times at which the wave function forms peaks).
For each location of the barrier ($c_0, c_1, ...$),
we measure the time between peaks of $| \psi(x_1,t) |^2$. At $c_0$ (a) the map
moves toward the center and then accumulates in the central region. The map
of $c_1$ (b) shows some ordered lines, and it is easy to see that
the wave packet stays most of the time in the left side. More ordered behavior
appears in the map of $c_2$ (c), where the lines appear as semi-periodic paths.
[The mapping oscillates back and forth, each time adding points to paths.]
Finally, $c_3$ (d) provides ordered maps, as expected. The same pattern can
be seen also in $ <\pi> : <\rho> $ ($\pi = d \rho / dt$), at peak times,
and from the Poincar\'e maps $ <p> : <x> $ sampled when $\rho(t)$ is minimum
(i.e. $\pi(t) = 0$ and $d \pi / dt > 0$) \cite{PatanPRL}.
A period doubling behavior (small circles within large circles)
can be seen in the $<p> : <x>$ plane, and in
the $ <\pi> : <\rho> $ plane, as shown in ref. \cite{Ash}.
The behavior becomes more
ordered, as the RE height decreases, while for $c_3$ there are just
large circles.

We have shown in this study that tunneling in the presence of almost
degeneracy of levels provides necessary and sufficient
 conditions for the development of
complex behavior of the wave functions of a quantum system.
 A large number of almost degenerate levels induces a high level of complexity.
In general a pair
of almost degenerate levels has a pair of equivalent eigenfunctions
for which one is non-alternating and the other alternating. Moreover,
in this case,
the probability to be in the left side is approximately equal to the
 probability
to be in the right side, without connection to the barrier position.
These conditions suggest an underlying symmetry property which depends
 strongly on the geometry of the system.
We wish to thank E. Eisenberg, M. Lewkowicz, I. Dana, and R. Berkovits for
many useful discussions.

\begin{figure}[tbp]
\caption{ (a) Probability to be in left side of the barrier
as a function of time for three
largest RE's (in the interval $(-l/5,l/5)$). (b) The power spectrum of (a).}
\label{fig.3}
\end{figure}

\begin{figure}[tbp]
\caption{ The entropy function as a function of time for different locations.
(a) $c=c_0=0$ (b) $c=c_1=-10.89$ (c) $c=c_2=-3.63$ (d) $c=c_3=-1$,
the inset shows an enlargement of typical periodic oscillations.}
\label{fig.4}
\end{figure}

\begin{figure}[tbp]
\caption{ The accumulation of points in the $<x>:<p>$ plane
according to peak times for different
locations. (a) $c=c_0=0$ (b) $c=c_1=-10.89$ (c) $c=c_2=-3.63$ (d) $c=c_3=-1$.
The doted line indicates the position of the center of the barrier.}
\label{fig.5}
\end{figure}

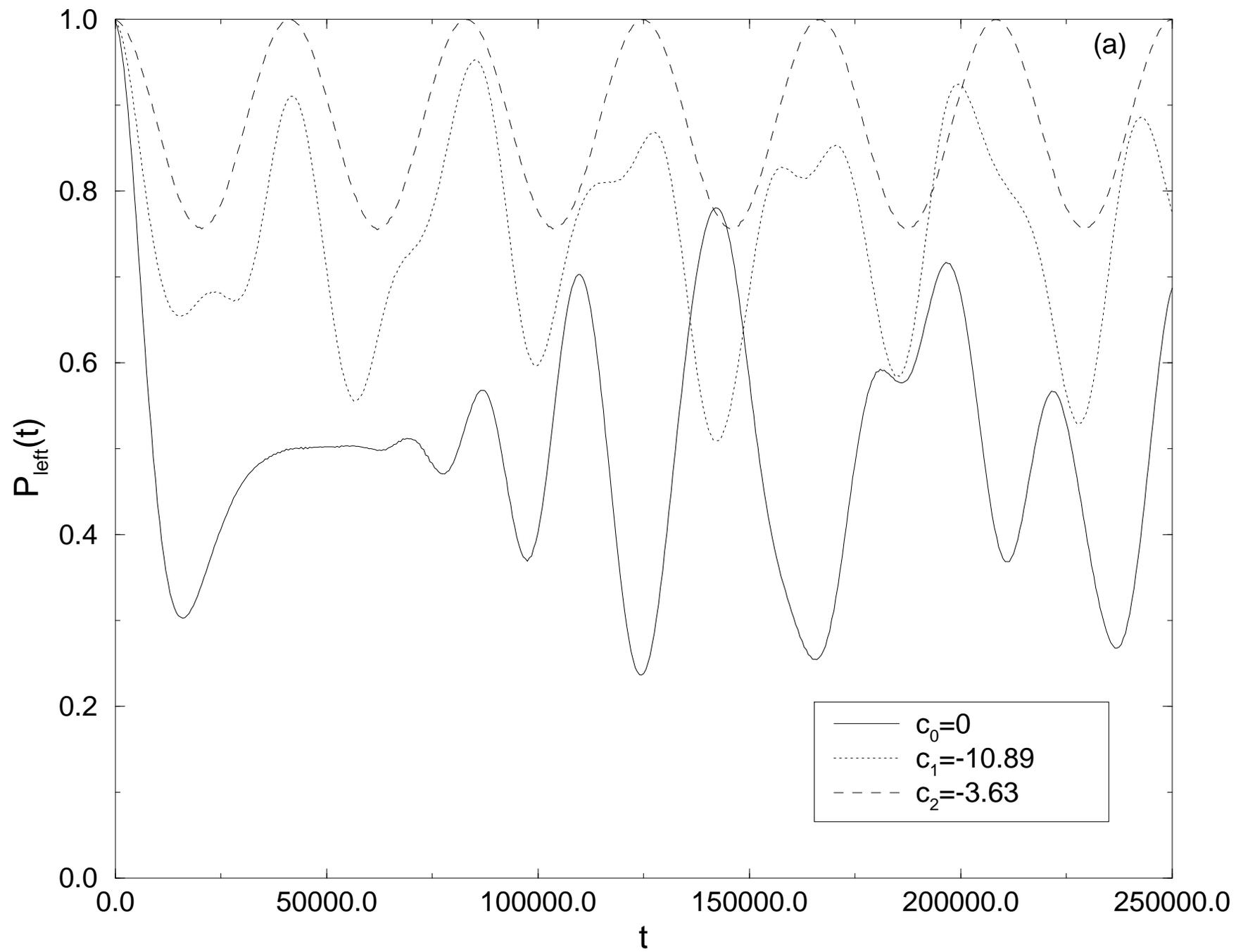
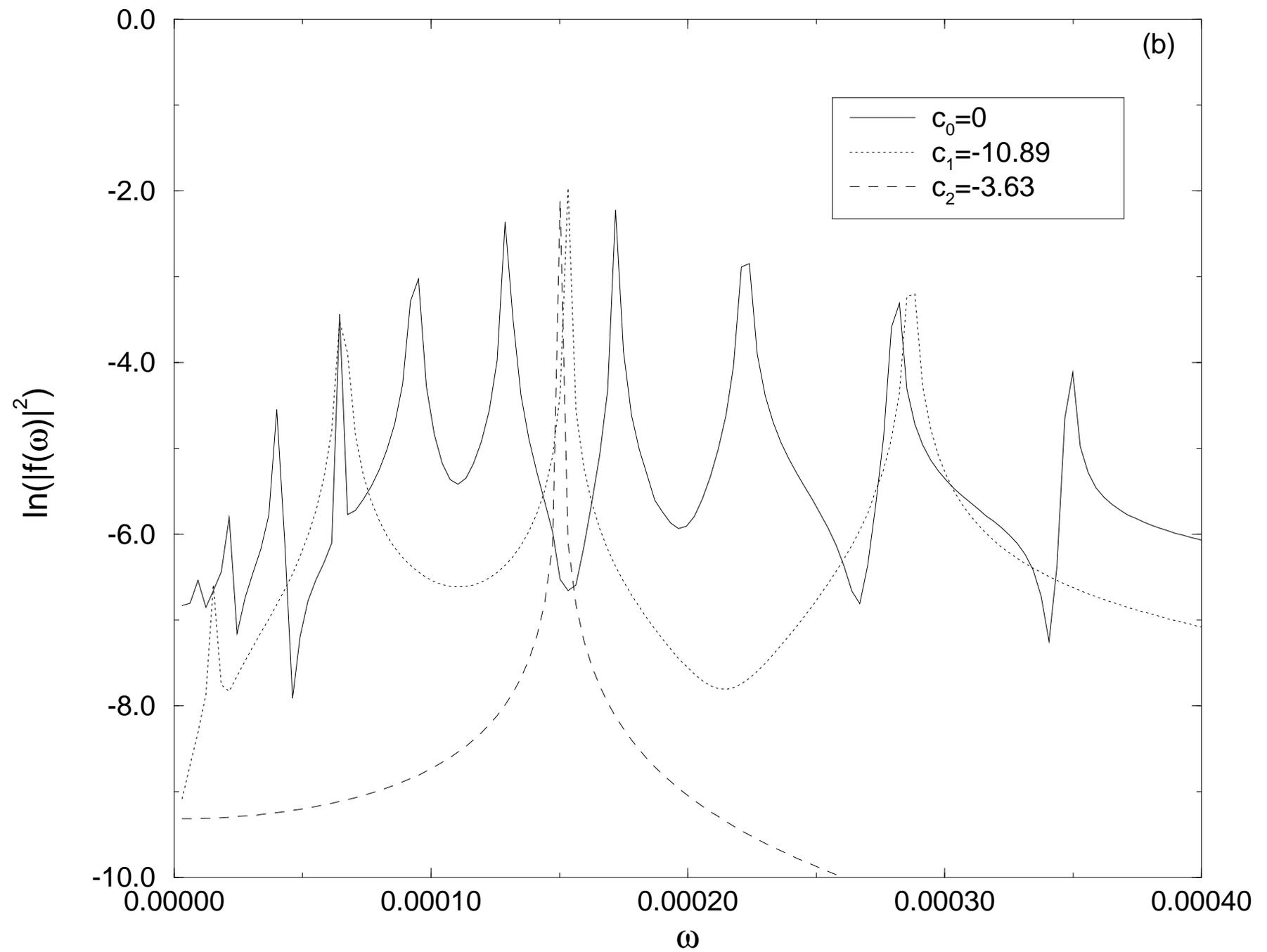
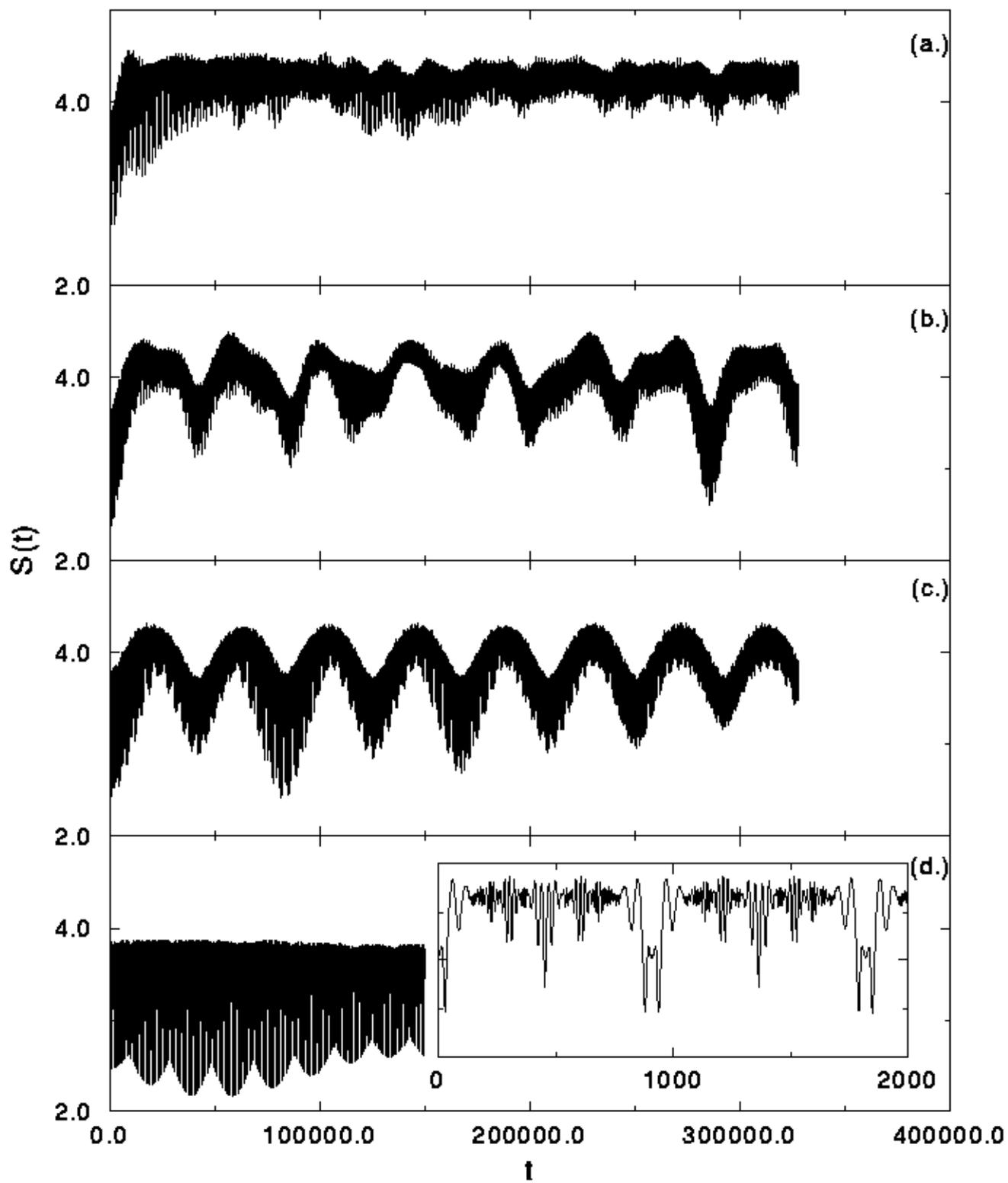
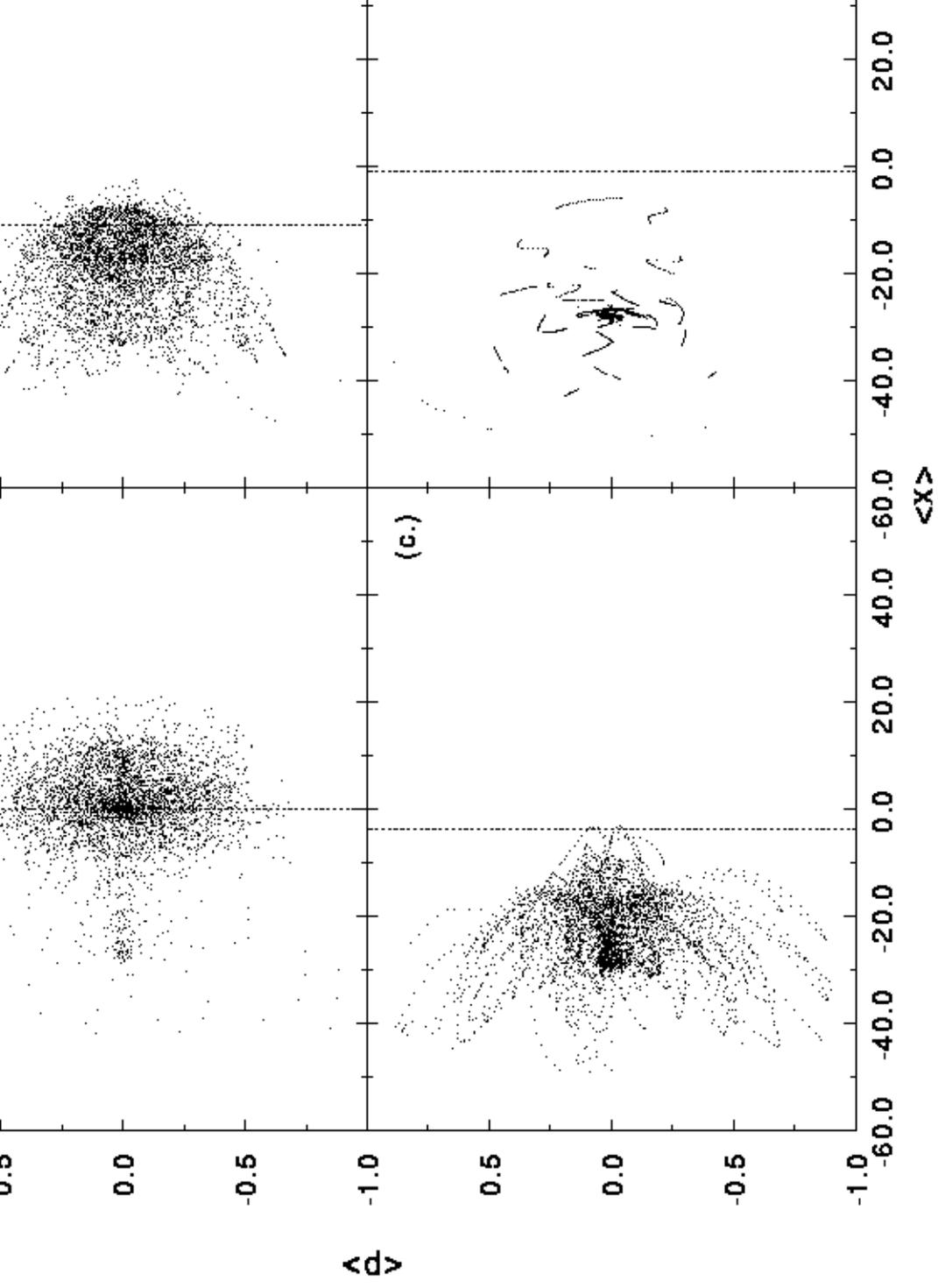
\end{document}